\documentclass{PoS}

\title{Astrophysical Implications of the QCD phase transition }

\ShortTitle{Astrophysical Implications of the QCD phase transition}

\author{\speaker{J\"urgen Schaffner-Bielich}$^a$,
Irina Sagert$^b$,
Matthias Hempel$^b$,
Giuseppe Pagliara$^a$\\
\llap{$^a$}Institut f\"ur Theoretische Physik, Ruprecht-Karls-Universit\"at, Philosophenweg 16,\\ D-69120 Heidelberg, Germany\\
\llap{$^b$}Institut f\"ur Theoretische Physik, Goethe Universit\"at, 
Max-von-Laue-Str.~1,\\ D-60438 Frankfurt am Main, Germany}

\author{Tobias Fischer$^c$,
Anthony Mezzacappa$^d$,
Friedrich-Karl Thielemann$^c$,
and Matthias~Liebend\"orfer$^c$\\
\llap{$^c$}Department of Physics, University of Basel, Klingelbergstr.~82, 4056 Basel, Switzerland\\
\llap{$^d$}Physics Division, Oak Ridge National Laboratory, Oak Ridge, TN~37831, USA}

\abstract{The possible role of a first order QCD phase transition at nonvanishing quark chemical potential and temperature for cold neutron stars and for supernovae is delineated. For cold neutron stars, we use the NJL model with nonvanishing color superconducting pairing gaps, which describes the phase transition to the 2SC and the CFL quark matter phases at high baryon densities. We demonstrate that these two phase transitions can both be present in the core of neutron stars and that they lead to the appearance of a third family of solution for compact stars. In particular, a core of CFL quark matter can be present in stable compact star configurations when slightly adjusting the vacuum pressure to the onset of the chiral phase transition from the hadronic model to the NJL model. We show that a strong first order phase transition can have strong impact on the dynamics of core collapse supernovae. If the QCD phase transition sets in shortly after the first bounce, a second outgoing shock wave can be generated which leads to an explosion. The presence of the QCD phase transition can be read off from the neutrino and antineutrino signal of the supernova.}

\FullConference{8th Conference Quark Confinement and the Hadron Spectrum\\
   September 1-6, 2008\\ Mainz. Germany}

\begin{document}

%%%%%%%%%%%%%%%%%%%%%%%%%%%%%%%%%%%%%%%%%%%%%%%%%%%%%%%%%%%%%%%%%%%%%%%%%%%

%\section{Introduction}

%%%%%%%%%%%%%%%%%%%%%%%%%%%%%%%%%%%%%%%%%%%%%%%%%%%%%%%%%%%%%%%%%%%%%%%%%%%

It has been realized in the last years, that the QCD phase diagram
might exhibit a rich structure at high baryon densities, be it in the
form of color superconducting phases \cite{Alford:2007xm} or the
quarkyonic phase \cite{McLerran:2007qj}. While recent lattice gauge
simulations indicate that the QCD phase transition at vanishing
quarkchemical potential is most likely a crossover, there might exist
a first order phase transition lines at nonvanishing quarkchemical
potentials. This region of the QCD phase diagram entails properties of
strongly interacting matter which are found in the core of neutron
stars and core-collapse supernovae and will be probed by heavy-ion
collisions with the CBM detector at the Facility for Antiproton and
Ion Research (FAIR) at GSI Darmstadt.

Neutron stars are born in core-collapse supernovae as so called proto-neutron stars. The temperatures reached in those supernovae and in proto-neutron stars are up to 50 MeV with baryon densities well above normal nuclear matter density. Similar conditions are encountered in simulations of neutron star mergers. 
The masses of rotation-powered neutron stars, pulsars, have been quite
accurately determined.  More than 1700 pulsars are presently known,
the best determined mass is the one of the Hulse-Taylor pulsar,
$M=(1.4414\pm 0.0002)M_\odot$ \cite{Weisberg:2004hi}, the smallest
known mass is $M=(1.18\pm 0.02)M_\odot$ for the pulsar J1756-2251
\cite{Faulkner:2005}. The most reliable lower limit for neutron star
masses published in the literature is the one of the Hulse-Taylor
pulsar (with only one noticeable exception
\cite{Champion:2008,Freire:2009}).  Note, that the mass of the pulsar
J0751+1807 was corrected from $M=2.1\pm 0.2 M_\odot$ down to
$M=1.14-1.40M_\odot$ \cite{Nice:2008}. 
The high masses and radii inferred from the X-ray burster EXO 0748--676 
in an analysis done in \cite{Ozel:2006km} are not model independent, a multiwavelength analysis 
concludes that a mass of $1.35M_\odot$ is more compatible with the data \cite{Pearson:2006zy}. In any case, high masses and radii do not exclude the possibility of having quark matter in the core of neutron stars
\cite{Alford:2006vz}.

%%%%%%%%%%%%%%%%%%%%%%%%%%%%%%%%%%%%%%%%%%%%%%%%%%%%%%%%%%%%%%%%%%%%%%%%%%%%

%\section{The QCD Phase Transition in Neutron Stars}

%%%%%%%%%%%%%%%%%%%%%%%%%%%%%%%%%%%%%%%%%%%%%%%%%%%%%%%%%%%%%%%%%%%%%%%%%%%%

In the following we will loosely denote the high-density matter as
quark matter, although the QCD phase transition at high baryon
densities is due to chiral symmetry breaking and not due to
deconfinement.  Let us first discuss the possible role of the QCD
phase transition and the stability of compact stars in a toy model for
quark matter with an equation of state of the form $p = a\cdot
\epsilon$ with a constant $a=1/3$ and a given energy density jump (see
\cite{Pagliara:2007ph}).  For the hadronic side, we use a relativistic
mean-field model fitted to the properties of nuclear matter (here set
GM3).  If the phase transition occurs close to the maximum mass,
always unstable solutions appear for the hybrid star with a quark
matter core. The mass-radius curve changes its slope as soon as quark
matter is present.  For an onset of the phase transition at moderate
densities, the presence of quark matter leads to stable
configurations, the slope of the mass-radius curve does not change its
sign \cite{Kaempfer81,Pagliara:2007ph}.

% \begin{figure}
% \centerline{
% \includegraphics[width=0.47\textwidth]{toy-eos1.epsi}
% \includegraphics[width=0.47\textwidth]{toy-mr1.epsi}}
% \caption{The equation of state (left plot) and the corresponding
%   mass-radius relation (right plot) of compact stars with a
%   first-order phase transition. For low critical densities, the hybrid
%   star configurations with a quark matter core are stable (taken from
%   \cite{Pagliara:2007ph})}.
% \label{fig:toy_mr}
% \end{figure}

Color-superconducting quark matter can be described by the NJL model
% \begin{eqnarray*}
% p &=& \frac{1}{2 \pi^2} \sum_{i=1}^{18} \int_0^\Lambda \mathrm{d} k \, k^2
% |\epsilon_i|+ 4 K \sigma_u \sigma_d \sigma_s
% - \frac{1}{4 G_D} \sum_{c=1}^{3} \left| \Delta_c \right|^2 \cr
% &&-2 G_S \sum_{\alpha=1}^{3} \sigma_\alpha^2+ \frac{1}{4
%   G_V}\omega_0^2+ p_e 
% \end{eqnarray*}
which includes both, dynamical quark masses via quark condensates and
the color-superconducting gaps $\Delta$ for the three flavor case (see
\cite{Ruster:2005jc,Blaschke:2005uj} for astrophysically relevant
calculations).  The parameters of the model are the cutoff, the scalar
and the vector coupling constants $G_S$, $G_V$, the diquark coupling
$G_D$, and the 't Hooft term coupling $K$. They are fixed to known
hadron masses, the pion decay constant, which leave two free
parameters, $G_D$ and $G_V$. In addition, the total pressure is
usually fixed by requiring that it vanishes in the vacuum. For the description of hybrid star matter, the results from the NJL model have to be merged with a low-density nuclear equation of state.
We demand that the
pressure constant is fixed such that the chiral phase transition
coincides with the transition from the hadronic model to the NJL model description. Numerically one finds, that the two different pressure constants differ only slightly from each other.
In the former case one finds a phase transition directly to CFL quark matter,
in the latter case two phase transitions appear, first to the 2SC phase then to the CFL phase.
% \begin{figure}
% \centerline{
% \includegraphics[width=0.5\textwidth]{mr-vec.epsi}}
% \caption{The mass-radius relation for neutron star with a color
%   superconducting quark matter core using the NJL model with a diquark
%   coupling constant $G_d=1.2G_s$ and a vector coupling constant
%   $G_v=0.2G_s$ (taken from \cite{Pagliara:2007ph}). The compact star
%   configurations with a quark matter core show stable branches of
%   solutions for the TOV equation.}
% \label{fig:mr_vec}
% \end{figure}
For a phase transition directly to the CFL phase, the solution is
first unstable but turns then into a stable one
\cite{Pagliara:2007ph}. The new stable solution is another example of
the third family of compact stars which can appear for a strong first
order phase transition
\cite{Gerlach68,Kaempfer81,Glendenning:1998ag,Schertler:2000xq}. For
the case of two phase transitions, two kinks appear in the mass-radius
curve with stable solutions for all configurations.  It is interesting
that it is possible that there are actually two phase transitions
present in compact star matter.

%%%%%%%%%%%%%%%%%%%%%%%%%%%%%%%%%%%%%%%%%%%%%%%%%%%%%%%%%%%%%%%%%%%%%%%%%%%

%\section{The QCD Phase Transition in Supernovae}

%%%%%%%%%%%%%%%%%%%%%%%%%%%%%%%%%%%%%%%%%%%%%%%%%%%%%%%%%%%%%%%%%%%%%%%%%%%%

The final state in the evolution of stars with a mass of more than 8
solar masses is a core-collapse supernova or a direct collapse to a black hole, see \cite{Janka:2006fh} for
an overview. The degenerate core collapses until normal nuclear matter
densities are reached. The repulsion between the nucleons halts the
collapse. A shock front is generated which moves outward but stalls
around 100 km. The newly born proto-neutron star emits neutrinos and
antineutrinos which deposit energy behind the stalled shock. For
low-mass progenitors with masses of eight to ten solar masses present
supernova simulations indeed find that this additional energy deposit
is enough to lead to a delayed explosion mechanism
\cite{Kitaura:2006}. For more massive progenitor stars, it was
recently proposed that the standing accretion shock instability (SASI)
is crucial for core-collapse supernova explosions after about
600ms \cite{Marek:2007gr}.

The conditions of core-collapse supernova matter at bounce are energy
densities of $\epsilon \sim (1-1.5)\epsilon_0$, so slightly above
normal nuclear matter energy density, temperatures of $T\sim 10-20$
MeV and proton fractions of $Y_p \sim 0.2-0.3$ (normal nuclear matter
has $Y_p=0.5$). In the following we explore the consequences of a QCD phase transition occurring shortly after the bounce in core-collapse supernovae \cite{Sagert:2008ka}. For simplicity, the MIT bag model is used to describe the high-density part of the equation of state, while for the low-density part the supernova equation of state of Shen et al. is adopted \cite{Shen98}. 
The mass-radius relation of neutron stars for an equation of state
with a QCD phase transition can be drastically different as shown
above.  The maximum masses are found to be $1.56 M_\odot$ for a bag
constant of $B^{1/4}=162$~MeV and $1.50 M_\odot$ for
$B^{1/4}=165$~MeV, so above the Hulse-Taylor pulsar mass limit.

% \centerline{
% \includegraphics[angle=-90,width=0.6\textwidth]{critdens.eps}}

The phase transition line of relevance for astrophysical applications
does not coincide with the one of ordinary nuclear matter as the
conditions are entirely different. The timescales are much longer so
that strangeness is not a conserved quantum number which makes quark
matter much more stable. Moreover, supernova matter is neutron-rich so
that nuclear matter is unfavoured due to the repulsive nature of the
asymmetry energy. Both effects lead to a considerably reduced critical
density for the QCD phase transition, so that the production of quark
matter in supernovae becomes more likely compared to the situation in
heavy-ion physics and can appear already close to the conditions at
bounce.

% \centerline{
% \includegraphics[angle=-90,width=0.5\textwidth]{critdens_B165-ud2.eps}}

For the situation in heavy-ion collisions, just up- and down-quark
matter has to be compared to symmetric nuclear matter, the timescales
in heavy-ion collisions are just too short to allow for net
strangeness being produced.  Large critical densities are found,
several times normal nuclear matter densities, so that the production
of ud-quark matter is unfavoured for heavy-ion collisions at small
temperatures and high baryon densities, there is no contradiction with
heavy-ion data. To be more specific, the freeze-out parameters
extracted from statistical models of particle production, $\mu_{f.o.}=
700-800$~MeV, $T_{f.o.}=50-70$~MeV for heavy-ion collisions at SIS
energies and $\mu_{f.o.} \sim 500$~MeV, $T_{f.o.}\sim 120$~MeV for AGS
energies, are well within the hadronic side of the phase diagram.

The equation of state with a QCD phase transition is used as input in a supernova simulation which also implements accurate neutrino transport of all neutrino flavours \cite{Sagert:2008ka}. 
Shortly after the bounce, the QCD phase transition is reached in the core of the proto-neutron star. The formation of a core of pure quark matter produces a second shock wave which is moving outwards leading to an explosion. Note, that in the standard 1D simulation without a phase transition no explosion is found, the first shock simply stalls and matter is eventually collapsing to a black hole.
The supernova simulation runs were performed for $10M_\odot$ and
$15M_\odot$ progenitor stars with two different bag parameters
controlling the onset of the QCD phase transition. We find that the
time of the appearance of the quark core, the baryonic
mass of the compact remnant and the explosion energy are significantly
sensitive to the location of the QCD phase transition.  With the curretn parameter sets the quark core
appears at $t_{\rm pb}=200$ to $500$~ms. If quark matter appears later, the explosion energy gets larger, as the density contrast at the accretion shock front increases. Heavy progenitor star masses can lead to the formation of a black hole. 
%
% \centerline{
% \includegraphics[height=\textheight]{sn_neutrinos_qgp.eps}}
%
There are observable implications for the QCD phase transition in the
neutrino signal of supernovae. When the second shock front produced by
the phase transition runs over the neutrinosphere a second burst of
anti-neutrinos is released. The peak location and height is determined
by the critical density and strength of the QCD phase transition.

%%%%%%%%%%%%%%%%%%%%%%%%%%%%%%%%%%%%%%%%%%%%%%%%%%%%%%%%%%%%%%%%%%%%%%%%%%%%

%\section{Summary}

%%%%%%%%%%%%%%%%%%%%%%%%%%%%%%%%%%%%%%%%%%%%%%%%%%%%%%%%%%%%%%%%%%%%%%%%%%%%

In summary, a first order QCD phase transition at high baryon density
can lead to observable astrophysical signals involving compact stars
and supernovae.  Neutron stars with a core of CFL quark matter can be
stable and can form a third family of compact stars
besides white dwarfs and ordinary neutron stars.  In an exploratory
study, we demonstrated that quark matter can be formed in
supernovae, even shortly after the bounce which produces a second
shock with enough energy to cause an explosion even in 1D
simulations.  The second shock forms a second peak in the
(anti-)neutrino signal. Possible implications for the gravitational
wave signal and r-process nucleosynthesis need to be explored in
future work. And much more refined models of QCD are needed to explore
the consequences of the QCD phase transitions at high baryon densities
in astrophysics on a more fundamental basis.

%\section*{Acknowledgments}

This work is supported by the German Research
Foundation (DFG) within the framework of the excellence initiative
through the Heidelberg Graduate School of Fundamental Physics,
the Gesellschaft f\"ur Schwerionenforschung
mbH Darmstadt, Germany, the Helmholtz Research School for Quark
Matter Studies, the
Helmholtz Alliance Program of the Helmholtz Association, contract HA-216 "Extremes of Density and Temperature: Cosmic Matter in the Laboratory",
the Frankfurt Institute for Advanced Studies,
the Italian National Institute for Nuclear Physics, the Swiss National
Science Foundation under the grant numbers PP002-106627/1 and
PP200020-105328/1, and the ESF CompStar program.
A.M. is supported at the Oak Ridge National Laboratory, managed by
UT-Battelle, LLC, for the U.S. Department of Energy under contract
DE-AC05-00OR22725.

\bibliographystyle{utphys}
\bibliography{all,literat,conf8refs}

\providecommand{\href}[2]{#2}\begingroup\raggedright\begin{thebibliography}{10}

\bibitem{Alford:2007xm}
M.~G. Alford, A.~Schmitt, K.~Rajagopal, and T.~{Sch\"afer}
  \href{http://arxiv.org/abs/0709.4635}{{\tt arXiv:0709.4635 [hep-ph]}}.

\bibitem{McLerran:2007qj}
L.~McLerran and R.~D. Pisarski {\em Nucl. Phys.} {\bf A796} (2007)  83--100,
  \href{http://arxiv.org/abs/0706.2191}{{\tt arXiv:0706.2191 [hep-ph]}}.

\bibitem{Weisberg:2004hi}
J.~M. Weisberg and J.~H. Taylor, ``The relativistic binary pulsar b1913+16:
  Thirty years of observations and analysis,'' in {\em Binary Radio Pulsars},
  F.~A. Rasio and I.~H. Stairs, eds., vol.~328 of {\em Astronomical Society of
  the Pacific Conference Series}, p.~25.
\newblock 2005.
\newblock \href{http://arxiv.org/abs/astro-ph/0407149}{{\tt astro-ph/0407149}}.

\bibitem{Faulkner:2005}
A.~J. {Faulkner}, M.~{Kramer}, A.~G. {Lyne}, R.~N. {Manchester}, M.~A.
  {McLaughlin}, I.~H. {Stairs}, G.~{Hobbs}, A.~{Possenti}, D.~R. {Lorimer},
  N.~{D'Amico}, F.~{Camilo}, and M.~{Burgay} {\em Astrophys. J.} {\bf 618}
  (2005)  L119--L122, \href{http://arxiv.org/abs/astro-ph/0411796}{{\tt
  astro-ph/0411796}}.

\bibitem{Champion:2008}
D.~J. {Champion}, S.~M. {Ransom}, P.~{Lazarus}, F.~{Camilo}, C.~{Bassa}, V.~M.
  {Kaspi}, D.~J. {Nice}, P.~C.~C. {Freire}, I.~H. {Stairs}, J.~{van Leeuwen},
  B.~W. {Stappers}, J.~M. {Cordes}, J.~W.~T. {Hessels}, D.~R. {Lorimer},
  Z.~{Arzoumanian}, D.~C. {Backer}, N.~D.~R. {Bhat}, S.~{Chatterjee},
  I.~{Cognard}, J.~S. {Deneva}, C.-A. {Faucher-Gigu{\`e}re}, B.~M. {Gaensler},
  J.~{Han}, F.~A. {Jenet}, L.~{Kasian}, V.~I. {Kondratiev}, M.~{Kramer},
  J.~{Lazio}, M.~A. {McLaughlin}, A.~{Venkataraman}, and W.~{Vlemmings} {\em
  Science} {\bf 320} (2008)  1309, \href{http://arxiv.org/abs/0805.2396}{{\tt
  arXiv:0805.2396 [astro-ph]}}.

\bibitem{Freire:2009}
{talk given at the International Conference on Dense Matter, Hirschegg,
  Austria, January 19-23, 2009}.

\bibitem{Nice:2008}
D.~J. {Nice}, I.~H. {Stairs}, and L.~E. {Kasian} {\em AIP Conference
  Proceedings} {\bf 983} (2008)  453--458.

\bibitem{Ozel:2006km}
F.~{\"O}zel {\em Nature} {\bf 441} (2006)  1115--1117,
  \href{http://arxiv.org/abs/astro-ph/0605106}{{\tt astro-ph/0605106}}.

\bibitem{Pearson:2006zy}
K.~J. Pearson, R.~Hynes, D.~Steeghs, P.~Jonker, C.~Haswell, A.~King,
  K.~O'Brien, G.~Nelemans, and M.~Mendez {\em Astrophys. J.} {\bf 648} (2006)
  1169--1180, \href{http://arxiv.org/abs/astro-ph/0605634}{{\tt
  astro-ph/0605634}}.

\bibitem{Alford:2006vz}
M.~Alford, D.~Blaschke, A.~Drago, T.~Kl\"ahn, G.~Pagliara, and
  J.~Schaffner-Bielich {\em Nature} {\bf 445} (2006)  E7--E8,
  \href{http://arxiv.org/abs/astro-ph/0606524}{{\tt astro-ph/0606524}}.

\bibitem{Pagliara:2007ph}
G.~Pagliara and J.~Schaffner-Bielich {\em Phys. Rev. D} {\bf 77} (2008)
  063004, \href{http://arxiv.org/abs/0711.1119}{{\tt arXiv:0711.1119
  [astro-ph]}}.

\bibitem{Kaempfer81}
B.~K{\"a}mpfer {\em J. Phys. A} {\bf 14} (1981)  L471--L475.

\bibitem{Ruster:2005jc}
S.~B. R{\"u}ster, V.~Werth, M.~Buballa, I.~A. Shovkovy, and D.~H. Rischke {\em
  Phys. Rev. D} {\bf 72} (2005)  034004,
  \href{http://arxiv.org/abs/hep-ph/0503184}{{\tt hep-ph/0503184}}.

\bibitem{Blaschke:2005uj}
D.~Blaschke, S.~Fredriksson, H.~Grigorian, A.~M. Oztas, and F.~Sandin {\em
  Phys. Rev. D} {\bf 72} (2005)  065020,
  \href{http://arxiv.org/abs/hep-ph/0503194}{{\tt arXiv:hep-ph/0503194}}.

\bibitem{Gerlach68}
U.~H. Gerlach {\em Phys. Rev.} {\bf 172} (1968)  1325.

\bibitem{Glendenning:1998ag}
N.~K. Glendenning and C.~Kettner {\em Astron. Astrophys.} {\bf 353} (2000)  L9,
  \href{http://arxiv.org/abs/astro-ph/9807155}{{\tt astro-ph/9807155}}.

\bibitem{Schertler:2000xq}
K.~Schertler, C.~Greiner, J.~Schaffner-Bielich, and M.~H. Thoma {\em Nucl.
  Phys.} {\bf A677} (2000)  463,
  \href{http://arxiv.org/abs/astro-ph/0001467}{{\tt astro-ph/0001467}}.

\bibitem{Janka:2006fh}
H.-T. Janka, K.~Langanke, A.~Marek, G.~Mart{\'i}nez-Pinedo, and B.~M{\"u}ller
  {\em Phys. Rept.} {\bf 442} (2007)  38--74,
  \href{http://arxiv.org/abs/astro-ph/0612072}{{\tt astro-ph/0612072}}.

\bibitem{Kitaura:2006}
F.~S. {Kitaura}, H.-T. {Janka}, and W.~{Hillebrandt} {\em Astron. Astrophys.}
  {\bf 450} (2006)  345--350,
  \href{http://arxiv.org/abs/arXiv:astro-ph/0512065}{{\tt
  arXiv:astro-ph/0512065}}.

\bibitem{Marek:2007gr}
A.~Marek and H.~T. Janka \href{http://arxiv.org/abs/0708.3372}{{\tt
  arXiv:0708.3372 [astro-ph]}}.

\bibitem{Sagert:2008ka}
I.~Sagert, M.~Hempel, G.~Pagliara, J.~Schaffner-Bielich, T.~Fischer,
  A.~Mezzacappa, F.-K. Thielemann, and M.~{Liebend\"orfer} {\em Phys. Rev.
  Lett.} {\bf 102} (2008)  081101, \href{http://arxiv.org/abs/0809.4225}{{\tt
  arXiv:0809.4225 [astro-ph]}}.

\bibitem{Shen98}
H.~Shen, H.~Toki, K.~Oyamatsu, and K.~Sumiyoshi {\em Nucl. Phys.} {\bf A637}
  (1998)  435--450, \href{http://arxiv.org/abs/nucl-th/9805035}{{\tt
  nucl-th/9805035}}.

\end{thebibliography}\endgroup

\end{document}